\documentclass[twocolumn,showpacs,preprintnumbers,prl,fleqn]{revtex4}
\usepackage{graphicx}
\begin{document}
\title{Interaction-driven spin precession in quantum-dot spin valves}
\author{J\"urgen K\"onig$^1$ and Jan Martinek$^{1,2}$}
\affiliation{
$^1$Institut f\"ur Theoretische Festk\"orperphysik, Universit\"at Karlsruhe,
76128 Karlsruhe, Germany\\
$^2$Institute of Molecular Physics, Polish Academy of Sciences, 60-179 
Pozna\'n, Poland}

\date{\today}

\begin{abstract}
We analyze spin-dependent transport through spin valves composed
of an interacting quantum dot coupled to two ferromagnetic leads.
The spin on the quantum dot and the linear conductance as a function of the 
relative angle $\theta$ of the leads' magnetization directions is derived to
lowest order in the dot-lead coupling strength.
Due to the applied bias voltage spin accumulates on the quantum dot,
which for finite charging energy experiences a torque, resulting in spin 
precession.
The latter leads to a non-trivial, interaction-dependent, $\theta$-dependence 
of the conductance.
In particular, we find that the spin-valve effect is reduced for all 
$\theta \neq \pi$.
\end{abstract}


\maketitle

{\it Introduction}. -- 
The field of spin- or magnetoelectronics has attracted much interest, 
for both its beautiful fundamental physics and its potential applications. 
One famous spin-dependent transport phenomenon is the tunnel 
magnetoresistance (TMR) in a {\em spin-valve} geometry, in which two 
ferromagnetic metals are separated by an insulating layer serving as a 
tunnel barrier \cite{julliere}. 
The transmission through the barrier decreases as the relative angle 
$\theta$ between the magnetizations of two ferromagnets is increased from
$0$ to $\pi$. 
Within a single-particle picture the $\theta$-dependent part of 
the transmission can be shown \cite{slonczewski,angular} to be proportional to 
$\cos \theta$.

Transport based on tunneling has also been extensively studied in
nanostructured devices such as semiconductor quantum dots (QDs) or
metallic single-electron transistors.
Recently, magnetotransport through those devices has attracted much interest. 
This includes normal or ferromagnetic metallic islands coupled to 
ferromagnetic leads \cite{ono,metal-theory} as well as spin-dependent 
transport from ferromagnets through QDs \cite{QD-theory,Kondo,QD-exp}.
Precession of a single magnetic atom spin in an external magnetic field has 
been detected, but only in the power spectrum of the tunneling 
current \cite{precession_e,precession_t}.

In this Letter, we study the effect of strong Coulomb interaction in a 
single-level QD (or a magnetic impurity \cite{impurity}) attached
to ferromagnetic leads on the average dot spin and the linear conductance
in the weak dot-lead coupling limit ($\Gamma \ll k_BT$, where $\Gamma$ is the
intrinsic line width of the dot levels) \cite{comment_0}.
We find an interaction-driven spin precession, even in the absence of an
external magnetic field.
This spin precession is predicted to be clearly visible in the linear 
conductance as a reduction of the spin-valve effect and a nontrivial 
$\theta$-dependence.
For any non-parallel configuration, transport is reduced 
as compared to the parallel one (spin-valve effect). 
In the absence of Coulomb interaction, the $\theta$-dependence follows 
simply $\cos \theta$ \cite{slonczewski,angular}. 
The presence of a finite charging energy, however, leads to a reduction of 
the spin-valve effect.
This can be understood by the interplay of spin accumulation caused by 
the bias voltage and an interaction-dependent spin torque due to
an effective exchange interaction between the spin in the dot and
the leads, which in turn generates spin precession, detectable in the
conductance.

\begin{figure}[h]
\centerline{\includegraphics[width=8.5cm]{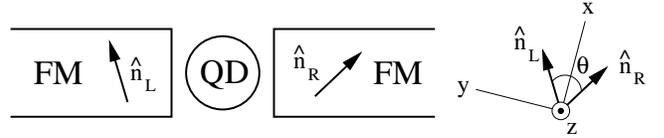}} 
\caption{
  A quantum-dot spin valve. 
  A quantum dot (QD) is connected to two ferromagnetic leads (FM). 
  The coordinate system we use is shown on the right.
  The magnetization directions (arrows) enclose an angle $\theta$.}
\label{fig1}
\end{figure}

{\it The model.} -- 
We consider a small QD with one spin-degenerate energy level 
$\epsilon$ participating in transport. 
The left and right lead are magnetized along $\hat \mathbf{n}_\mathrm{L}$ and 
$\hat \mathbf{n}_\mathrm{R}$ (see Fig.~\ref{fig1}), with relative angle 
$\theta$. 
The total Hamiltonian is $H = H_\mathrm{dot} +
H_\mathrm{L} + H_\mathrm{R} + H_\mathrm{T,L} + H_\mathrm{T,R}$.
The first part, $ H_\mathrm{dot} = \epsilon \sum_\sigma
c^\dagger_\sigma c_\sigma + U n_\uparrow n_\downarrow$, describes
the dot energy level plus the charging energy $U$ for double
occupation of the dot.
The leads are modeled by $H_r = \sum_k \epsilon_k a^\dagger_{rk}a_{rk}$ with 
$r = \mathrm{L,R}$.
For simplicity, we assume them to be half metallic, i.e., only 
majority spins have a finite density of states. 
Tunneling between lead and dot is described by 
$H_\mathrm{T,L} = t \sum_k \left( a^\dagger_{\mathrm{L}k} c_+ + h.c. \right)$, 
where $c_+$ is the Fermi operator for an electron on the QD with spin 
along $\hat {\bf n}_{\rm L}$.
It is convenient to quantize the dot spin along the $z$-direction in the 
coordinate system defined by 
$\hat \mathbf{e}_x = (\hat \mathbf{n}_\mathrm{L}+ \hat \mathbf{n}_\mathrm{R}) 
/ |\hat \mathbf{n}_\mathrm{L}+\hat \mathbf{n}_\mathrm{R}|$, 
$\hat \mathbf{e}_y = (\hat \mathbf{n}_\mathrm{L}- \hat \mathbf{n}_\mathrm{R}) 
/ |\hat \mathbf{n}_\mathrm{L}-\hat \mathbf{n}_\mathrm{R}|$, 
and 
$\hat \mathbf{e}_z = (\hat \mathbf{n}_\mathrm{R} \times 
\hat \mathbf{n}_\mathrm{L})/ |\hat \mathbf{n}_\mathrm{R} \times 
\hat \mathbf{n}_\mathrm{L}|$.
The tunnel Hamiltonian, then, is
\begin{equation}
  H_\mathrm{T,L} = {t\over \sqrt{2}} \sum_k
  \left( e^{i\theta/4} a^\dagger_{\mathrm{L}k} c_\uparrow
    + e^{-i\theta/4} a^\dagger_{\mathrm{L}k} c_\downarrow
  + h.c. \right)
  \, ,
\label{tunelling}
\end{equation}
and $H_\mathrm{T,R}$ is the same but with $\mathrm{L} \rightarrow \mathrm{R}$ 
and $\theta \rightarrow - \theta$.
Due to tunneling the dot level acquires a finite width
$\Gamma = 2\pi |t|^2 N$, where $N = N_\mathrm{L}=N_\mathrm{R}$ is the density
of states of the majority spins of the left and right lead.

With this choice of the quantization axis, the model 
studied here appears similar to those for Aharonov-Bohm interferometers 
which contain a single-level QD in each arm \cite{AB1,AB2,AB3}. 
In the present case, the two ``arms'' are labeled by the spin 
$\sigma = \uparrow,\downarrow$ along the $z$-direction, and the 
Aharonov-Bohm phase due to an enclosed magnetic flux corresponds to the 
angle $\theta$ between the leads' magnetizations. 
The limit $U\rightarrow \infty$ of the present model is, in fact, equivalent 
to the one studied in Sec. IV. C. of Ref.~\onlinecite{AB2}.

{\it Linear conductance and average spin.} -- 
We make use of the analogy between the quantum-dot spin valve and the
Aharonov-Bohm interferometer and express the current in terms of
Green's functions of the dot electrons, as shown in Eq.~(4.3) of
Ref.~\onlinecite{AB2}. 
Here, we are interested in first-order transport in $\Gamma$, 
for which the linear-response conductance 
$G ^\mathrm{lin} = (\partial I / \partial V)\big|_{V=0}$ simplifies to 
\cite{comment_1}
\begin{eqnarray}
   G^\mathrm{lin} &=&
        {e^2 \over h} \Gamma \int d \omega \, \left\{
        {\rm Im} \, G_{\downarrow\downarrow}^{\mathrm{ret}}(\omega) f' (\omega)
\right. \nonumber \\
        &&\left.
        + \sin{\theta \over 2} f(\omega) \,
        {\partial G^>_{\downarrow\uparrow}(\omega) \over \partial (eV)}
\right. \nonumber \\
        &&\left.
        + \sin{\theta \over 2}
        \left[ 1 - f(\omega) \right]
        {\partial G^<_{\downarrow\uparrow}(\omega) \over \partial (eV)}
        \right\} \, .
\label{Greens}
\end{eqnarray}
Here, $f(\omega)$ is the Fermi function, $G_{\sigma\sigma'}(\omega)$
are the Fourier transforms of the Green's functions 
$G^>_{\sigma\sigma'}(t) = -i \langle c_\sigma (t) c_{\sigma'}^\dagger(0) 
\rangle$, $G^<_{\sigma\sigma'}(t) = i \langle c_{\sigma'}^\dagger(0) 
c_\sigma (t) \rangle$, and $G^\mathrm{ret}_{\sigma\sigma'}$ is the usual 
retarded Green's function.
Contributions involving 
${\rm Im} \, G_{\uparrow\uparrow}^{\mathrm{ret}}(\omega)$,
$G^>_{\uparrow\downarrow}(\omega)$, and
$G^<_{\uparrow\downarrow}(\omega)$ are accounted for in the prefactor 2.
Since $\Gamma$ already appears explicitly in front of the integral, 
all Green's functions are to be taken to zeroth order in $\Gamma$. 
In this limit, we find $-(1/ \pi) \mathrm{Im} \,
G^{\mathrm{ret}}_{\downarrow\downarrow} (\omega) = 
( P^0_0 + P^\downarrow_\downarrow ) \delta(\omega - \epsilon) + 
(P^\uparrow_\uparrow + P^d_d ) \delta(\omega - \epsilon - U)$,
$G^>_{\downarrow\uparrow} (\omega) = 2\pi i P^\downarrow_\uparrow
\delta(\omega - \epsilon - U)$, and 
$G^<_{\downarrow\uparrow} (\omega) = 2\pi i P^\downarrow_\uparrow 
\delta(\omega - \epsilon)$, where 
$P^\chi_{\chi'} = \langle |\chi'\rangle\langle \chi| \rangle$ are elements of 
the stationary density matrix (to zeroth order in $\Gamma$) of the quantum 
dot subsystem, with $\chi,\chi' = 0$ (empty dot), $\uparrow, \downarrow$
(singly-occupied dot), and $d$ (doubly-occupied dot).

The main task is now to determine these density matrix elements to zeroth 
order in $\Gamma$. 
They contain as well the information about the average occupation and spin on 
the QD. 
The diagonal matrix elements, $P^\chi_{\chi}$, are nothing but the
probabilities to find the QD in state $\chi$, i.e., the
dot is empty with probability $P_0^0$, singly occupied with $P_1
\equiv P_\uparrow^\uparrow + P_\downarrow^\downarrow$, and doubly
occupied with $P_d^d$. 
A finite spin can only emerge for single occupancy. 
The average spin $\mathbf{S} = (S_x,S_y,S_z)$ is related to 
the off-diagonal matrix element $P^\downarrow_\uparrow$ via 
$S_x = \mathrm{Re} \, P^\downarrow_\uparrow$, 
$S_y = \mathrm{Im} \, P^\downarrow_\uparrow$, 
and $S_z = (1/2) (P^\uparrow_\uparrow - P^\downarrow_\downarrow)$.

It is remarkable that on the r.h.s of Eq.~(\ref{Greens}) derivatives of 
Green's function with respect to bias voltage $V$ appear. 
As a consequence, not only the equilibrium density matrix elements enter the 
linear conductance, but also linear corrections in $V$ are involved. 
In equilibrium, $V=0$, the density matrix is diagonal with $P^0_0 = 1/ Z$, 
$P^\uparrow_\uparrow = P^\downarrow_\downarrow = \exp(-\beta \epsilon) / Z$, 
$P^d_d = \exp[-\beta (2\epsilon+U)] / Z$, with
$Z = 1+ 2\exp(-\beta \epsilon)+\exp[-\beta (2\epsilon+U)]$ and 
$\beta = 1/(k_\mathrm{B}T)$. 
As a consequence, the average spin on the QD vanishes at $V=0$ 
\cite{comment_2}.
With applied bias voltage, though, a finite spin can accumulate, which yields 
a finite $(d\mathbf{S}/dV)\big|_{V=0}$.

We now determine the needed density matrix elements by using the
real-time transport theory developed in Ref.~\onlinecite{diagrams}. 
The starting point is the generalized stationary master equation in Liouville 
space,
\begin{equation}
\label{master general}
   \left( \epsilon_{\chi_1} - \epsilon_{\chi_2} \right) P^{\chi_1}_{\chi_2}
        + \sum_{\chi_1',\chi_2'} P^{\chi_1'}_{\chi_2'}
        \Sigma^{\chi_1',\chi_1}_{\chi_2',\chi_2} = 0 \, ,
\end{equation}
where $\chi_1$ and $\chi_2$ label the QD states, and $\epsilon_{\chi_1}$ and 
$\epsilon_{\chi_2}$ are the corresponding energies. 
The matrix elements $P^{\chi_1}_{\chi_2}$ of the density
matrix are connected to each other in Eq.~(\ref{master general})
by terms $\Sigma^{\chi_1',\chi_1}_{\chi_2',\chi_2}$ which can
be viewed as generalized transition rates in Liouville space. 
They are defined as irreducible self-energy parts of the propagation in 
Liouville space and are represented as diagram blocks on a Keldysh contour. 
For a detailed derivation of this diagrammatic language, the generalized 
master equation, and the rules on how to calculate a diagram we refer 
to Ref.~\onlinecite{diagrams}.

In the following, we write $P=\bar P + \hat P +\ldots$ and $\Sigma
= \bar \Sigma + \hat \Sigma + \ldots$, where $\bar P$ and $\bar
\Sigma$ denote the equilibrium limit, and $\hat P$ and $\hat
\Sigma$ are the linear corrections in $V$.
Using the symmetry of the model, we find the relations 
$\bar P_\chi^{\chi'} = \bar P_{\tilde \chi}^{\tilde \chi'}$ and 
$\bar \Sigma_{\chi,\chi''}^{\chi',\chi'''} = \bar
\Sigma_{\tilde \chi, \tilde \chi''}^{\tilde \chi',\tilde \chi'''}$
and $\hat P_\chi^{\chi'} = - \hat P_{\tilde \chi}^{\tilde \chi'}$
and $\hat \Sigma_{\chi,\chi''}^{\chi',\chi'''} = - \hat
\Sigma_{\tilde \chi, \tilde \chi''}^{\tilde \chi',\tilde \chi'''}$, where 
$\tilde \chi$ is obtained from $\chi$ by the transformation 
$\uparrow \leftrightarrow \downarrow$.
For transitions from diagonal states in Liouville space to diagonal ones we 
find $\hat \Sigma_{\chi,\chi'}^{\chi,\chi'}=0$.
Finally, we drop all $\Sigma$ terms which connect states in Liouville space 
that are not compatible, at least to lowest order in $\Gamma$. 
It turns out that it is sufficient to specify Eq.~(\ref{master general}) for
$\chi_1 = \chi_2 = \downarrow$ as well as for $\chi_1 =
\downarrow$, $\chi_2 = \uparrow$. 
For the linear correction in $V$ we get
\begin{eqnarray}
\label{master 1}
  0 &=& \hat P_{\downarrow}^{\downarrow}
  \bar \Sigma_{\downarrow,\downarrow}^{\downarrow,\downarrow}
  + \hat P_{\uparrow}^{\downarrow} \left(
    \bar \Sigma_{\uparrow,\downarrow}^{\downarrow,\downarrow}
    - \bar \Sigma_{\downarrow,\downarrow}^{\uparrow,\downarrow} \right)
\\
 0 &=& \hat P_{\uparrow}^{\downarrow}
 \bar \Sigma_{\uparrow,\uparrow}^{\downarrow,\downarrow}
 + \hat P_{\downarrow}^{\downarrow} \left(
   \bar \Sigma_{\downarrow,\uparrow}^{\downarrow,\downarrow}
   - \bar \Sigma_{\uparrow,\uparrow}^{\uparrow,\downarrow} \right)
\nonumber \\
        &&+ \bar P_0^0 \hat \Sigma_{0,\uparrow}^{0,\downarrow}
        + \bar P_{\downarrow}^{\downarrow} \left(
        \hat \Sigma_{\downarrow,\uparrow}^{\downarrow,\downarrow}
        + \hat \Sigma_{\uparrow,\uparrow}^{\uparrow,\downarrow} \right)
        + \bar P_{d}^{d} \hat \Sigma_{d,\uparrow}^{d,\downarrow} \, .
\label{master 2}
\end{eqnarray}
We evaluate the all necessary diagrams 
$\Sigma^{\chi_1',\chi_1}_{\chi_2',\chi_2}$ explicitly.
Eventually, we find the solution
\begin{equation}
  \hat P_{\uparrow}^{\downarrow} = {i\over 4} {eV\over k_B T}
  \bar P_1 \cos^2 \alpha(\theta) \sin {\theta\over 2}
\end{equation}
and $\hat P_{\downarrow}^{\downarrow} = -\hat P_{\uparrow}^{\uparrow}
= i \hat P_{\uparrow}^{\downarrow} \tan \alpha(\theta)$.
We used the definition
\begin{equation}
   \tan \alpha(\theta)
 = {A \over 1-f(\epsilon)+f(\epsilon+U)} \cos{\theta \over 2}\, ,
\label{angle}
\end{equation}
where $A = {1\over \pi} \mathrm{Re} \left[
\Psi\left({1\over 2}+i{\beta\epsilon\over 2\pi}\right)
-\Psi\left({1\over 2}+i{\beta(\epsilon+U)\over 2\pi}\right) \right]$, and 
$\Psi(x)$ denotes the digamma function.
This means that the spin accumulated in the QD is
(for $eV \ll k_BT$)
\begin{equation}
  |\mathbf{S}| =\sqrt{S_y^2+S_z^2} = {eV\over 4k_B T} \bar P_1 \cos
\alpha(\theta) \sin {\theta \over 2} \, ,
 \end{equation}
with $\alpha(\theta)$ being the angle enclosed by the quantum-dot
spin and the $y$-axis, $\tan \alpha(\theta)=S_z/S_y$.

With the result for $\hat P_\uparrow^\downarrow$ we are able to obtain the
linear conductance.
It can be written in the compact form
\begin{equation}
   G^\mathrm{lin} = G^\mathrm{lin, max}
   \left( 1 - \cos^2 \alpha(\theta) \sin^2{\theta \over 2} \right) \, .
\end{equation}
This equation together with the condition for $\alpha(\theta)$,
Eq.~(\ref{angle}), is the central result of this paper. 
The conductance is maximal for parallel magnetization, $\theta=0$. 
Its value is $G^\mathrm{lin, max} = (\pi e^2/ h) (\Gamma/ k_BT)
[1-f(\epsilon+U)]f(\epsilon) [1-f(\epsilon)+f(\epsilon+U)] /
[f(\epsilon)+1-f(\epsilon+U)]$.

We can straightforwardly generalize our theory to allow for arbitrary spin 
polarization 
$p \equiv (N_\mathrm{maj} - N_\mathrm{min})/(N_\mathrm{maj} + N_\mathrm{min})$
in the leads.
In this case, we get
$G^\mathrm{lin}/G^\mathrm{lin, max} =  1 - p^2 \cos^2 \alpha_p(\theta) 
\sin^2(\theta/2)$ with $\tan \alpha_p(\theta) = p \tan \alpha_{p=1}(\theta)$.

A reduction of the spin-valve effect has also been found for hybrid systems 
of ferromagnets with Luttinger liquids \cite{egger} or normal metals
\cite{brataas}.
Both its physical origin and the $\theta$-dependence of the conductance are
different from our proposal, though.

{\it Results and discussion.} -- 
It is interesting to analyze how Coulomb interaction in the dot affects the 
spin accumulation in our model.
In the absence of charging energy, $U=0$, we find $A=0$ and $\alpha =0$, and
the accumulated spin is along the $y$-direction, i.e., along
$\hat \mathbf{n}_L - \hat \mathbf{n}_R$.
A finite charging energy, however, yields a rotation of the quantum-dot 
spin within the $y-z$-plane by an angle $\alpha$, accompanied by a reduction 
of the total accumulated spin by $\cos\alpha$. 
The origin of the torque responsible for this spin rotation is an 
interaction-dependent effective spin exchange of the quantum-dot spin with 
the lead spins.
In the subspace of single dot occupation an effective spin Hamiltonian can 
be derived from the full model by means of a Schrieffer-Wolff transformation. 
If we now employ a mean-field picture and replace the lead spin operators by 
their average value, we end up with the simple effective Hamiltonian 
$H^\mathrm{eff} = A\Gamma \cos(\theta/2) S_x$ for the subspace under 
consideration.
In this effective model, a spin in the $y-z$-plane experiences a torque and
starts to precess, as described by classical Bloch equations. 
Together with the rate $\Gamma_{1\rightarrow 0} +\Gamma_{1\rightarrow d} =
\Gamma [1-f(\epsilon)+f(\epsilon+U)]$ to diminish the $z$-component of the 
quantum-dot spin by changing the dot state from single occupation to an empty
or doubly-occupied dot via tunneling, we get the Bloch equation 
$dS_z/dt = A\Gamma \cos(\theta/2) S_y - (\Gamma_{1\rightarrow 0}
+\Gamma_{1\rightarrow d}) S_z$, 
from which we can extract the angle $\alpha(\theta)$ of the 
rotated spin in the stationary limit $dS_z/ dt =0$, and we recover 
Eq.~(\ref{angle}).

We emphasize that both the rates for changing the number of dot electrons and
the spin precession are of the same order in $\Gamma$, which is why the 
the angle $\alpha$ is $\Gamma$-independent.
The two types of processes correspond to two different kinds of diagrams 
$\Sigma^{\chi_1',\chi_1}_{\chi_2',\chi_2}$.
In spin-precession terms, all four labels $\chi_1,\chi_1',\chi_2',\chi_2$ 
represent single occupation, $\uparrow,\downarrow$.
They describe first-order virtual charge fluctuations during which the spin 
is rotated.
In contrast, tunneling rates which change the number of dot electrons are
described by diagrams with $\chi_1 = \chi_1' = 0$ or 
$\chi_1 = \chi_1' = d$ or $\chi_2 = \chi_2' = 0$, or 
$\chi_2 = \chi_2' = d$.
It is crucial for a consistent theory of first-order transport to include 
both types of diagrams in Eqs.~(\ref{master 1}) and (\ref{master 2}).

\begin{figure}[h]
\centerline{\includegraphics[width=8cm]{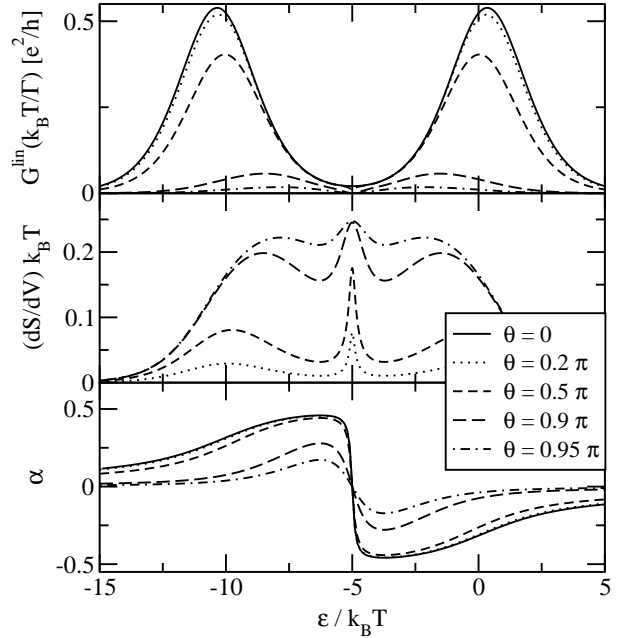}} \caption{Upper
panel: Linear conductance (normalized by $\Gamma/k_BT$ and plotted
in units of $e^2/h$) as a function of level position $\epsilon$
for five different angles $\theta$. Middle panel: Derivative of
accumulated spin $S$ with respect to bias voltage $V$ normalized
by $k_BT$. Lower panel: angle $\alpha$ between the quantum-dot
spin and the $y$-axis. In all panels we have chosen the charging
energy $U/k_BT = 10$ and half-metallic leads. } 
\label{fig2}
\end{figure}

The linear conductance as a function of the level energy $\epsilon$ is 
plotted in the upper panel of Fig.~\ref{fig2} for interaction strength 
$U/k_BT=10$ and different values of the angle $\theta$. 
For parallel magnetization, $\theta=0$, there are two conductance peaks 
located near $\epsilon=0$ and $\epsilon = -U$, respectively. 
With increasing angle $\theta$, transport is more and more suppressed
due to the spin-valve effect. 
However, this suppression is not uniform, as would be in the absence of 
charging energy. 
In contrast, the spin-valve effect is less pronounced in the valley
between the two peaks, where the dot is dominantly singly occupied,
and spin accumulation can occur. 
As a consequence, the two peaks move towards each other with increasing 
$\theta$.

The differential spin accumulation $dS/dV$ in units of $k_BT$ is illustrated
in the middle panel of Fig.~\ref{fig2}.
It is clear that single occupation of the dot is required for spin
accumulation, i.e., the plotted signal is high in the valley between the
two conductance peaks.

As explained above, an effective exchange interaction between
quantum-dot spin and spin of the leads yield a rotation of the
accumulated spin in the $y-z$-plane by an angle $\alpha$. 
The lower panel of Fig.~\ref{fig2} depicts the evolution of $\alpha$
as a function of the level energy $\epsilon$. 
This angle is large in the valley between the conductance peaks, getting 
close to $\pm \pi/2$. 
A special point is $\epsilon = -U/2$, at which, due to particle-hole 
symmetry, the effective exchange interaction vanishes.
As a consequence, $\alpha$ shows a sharp transition from positive to negative 
values, accompanied with a peak in the accumulated spin.

\begin{figure}[h]
\centerline{\includegraphics[width=8cm]{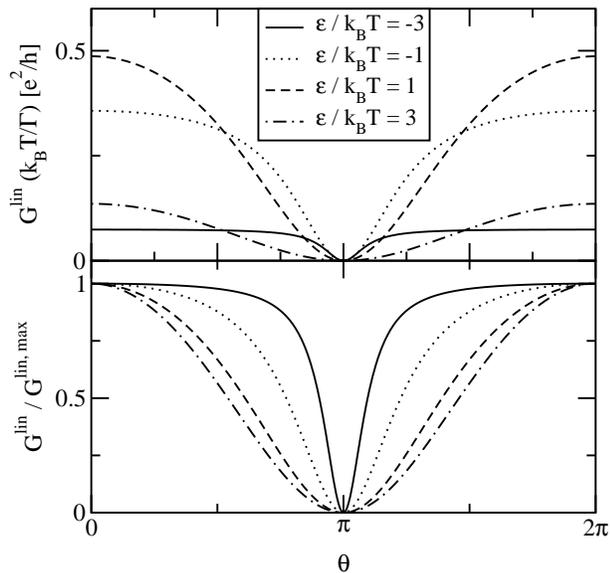}} 
\caption{Upper panel: Linear conductance as a function of $\theta$ for 
  $U/k_BT = 10$ and four different values of the level position. 
  Lower panel: The same but normalized to the maximal conductance for
  parallel magnetization.} 
\label{fig3}
\end{figure}

In the upper panel of Fig.~\ref{fig3} we show the linear conductance as a 
function of $\theta$ for four values of $\epsilon$. 
For $\epsilon/k_BT = 3$ and $1$, the dot is predominantly empty, and the 
$\theta$-dependence of the conductance is almost harmonic. 
For $\epsilon/k_BT = -1$ and $-3$, however, the spin-valve effect is strongly 
reduced, and conductance is enhanced, except in the regime close to 
antiparallel magnetization, $\theta=\pi$.
The enhancement of conductance is related to the fact that the spin precession
reduces the angle between the accumulated spin and the magnetization direction 
of the drain electrode.
This is even better illustrated in the lower panel of
Fig.~\ref{fig3} which shows the same curves but normalized to
conductance at parallel magnetization ($\theta=0$). 
For $\epsilon/k_BT = -3$, the conductance stays almost flat over a
broad range, and then establishes the spin-valve effect only in a
small region around $\theta=\pi$.

Finally, we comment that a finite spin-flip relaxation time $\tau_\mathrm{sf}$
will reduce the spin-valve effect and limit its observability to 
$\Gamma > \tau_\mathrm{sf}^{-1}$ \cite{fujisawa}.
The main prediction of our theory, the deviation from the 
$\cos \theta$-law, will not be affected by $\tau_\mathrm{sf}$, as long as
a $\theta$-dependence is visible.

To summarize, there are two pronounced features in the linear conductance 
which proves the spin precession proposed in this Letter:
(i) the shift of two adjacent conductance peaks towards each other with
increasing angle $\theta$ (Fig.~\ref{fig2}), and (ii) strong deviation from 
the $\cos\theta$ law for the spin-valve effect (Fig.~\ref{fig3}).

{\it Acknowledgements.} --
We thank J. Barna{\'s}, A. Brataas, M. Braun, L. Glazman, S. Maekawa, and 
G. Sch\"on for discussions.
This work was supported by the Deutsche Forschungsgemeinschaft under the 
Center for Functional Nanostructures and the Emmy-Noether program.


\begin{thebibliography}{99}
\bibitem{comment_0}
In this limit the Kondo effect discussed in Ref.~\onlinecite{Kondo} does not
appear.
\bibitem{comment_1}
This equation corresponds to Eq.~(4.4) of Ref.~\onlinecite{AB2} but 
with the dot labels $1$ and $2$ replaced by $\downarrow$ and $\uparrow$, and 
with no additional spin index to be summed over.
\bibitem{comment_2}
In equilibrium and for lowest order, the finite magnetization in the 
leads does not yield a spin accumulation in the dot. 
To understand this let us first include a finite density of states for 
minority spins in the leads. Then, majority-spin electrons enter the 
QD at a higher rate than minority spins but this is exactly 
compensated for by a higher rate for the majority spin to leave the dot. 
This remains true also in the half-metallic limit.
\end{thebibliography}
\end{document}